\documentclass[a4paper,11pt]{article}
\usepackage[dvips]{graphicx}
\usepackage{amssymb}
\usepackage{amsmath}

\usepackage{cite}

\begin{document}

\title{Extended Supersymmetric Structures in Gapped and Superconducting Graphene}
\author{
V.K. Oikonomou$^{1,}$\,\thanks{v.k.oikonomou1979@gmail.com,voiko@physics.auth.gr}\\
$^1$Department of Theoretical Physics, Aristotle University of Thessaloniki,\\
54124 Thessaloniki, Greece
} \maketitle

\begin{abstract}
In view of the many quantum field theoretical descriptions of graphene in $2+1$ dimensions, we present another field theoretical feature of graphene, in the presence of defects. Particularly, we shall be interested in gapped graphene in the presence of a domain wall and also for superconducting graphene in the presence of a vortex. As we explicitly demonstrate, the gapped graphene electrons that are localized on the domain wall are associated with four $N=2$ one dimensional supersymmetries, with each pair combining to form an extended $N=4$ supersymmetry with non-trivial topological charges. The case of superconducting graphene is more involved, with the electrons localized on the vortex being associated with $n$ one dimensional supersymmetries, which in turn combine to form an $N=2n$ extended supersymmetry with no-trivial topological charges. As we shall prove, all supersymmetries are unbroken, a feature closely related to the number of the localized fermions and also to the exact form of the associated operators. In addition, the corresponding Witten index is invariant under compact and odd perturbations. 
\end{abstract}

\section*{Introduction}

Graphene is one of the most promising physical systems discovered in the last ten years, since it's properties offer place for many physical theories to be studied and observed in the laboratory. Particularly, graphene can be modelled by using relativistic quantum mechanics and quantum field theoretic methods \cite{graph1,graph2,graph3,graph4,graph5,graph6,graph7,graphfourfermion} giving rise to interesting and useful properties for electrons \cite{graphnew1,wilzek,graphnew3,graphnew4,graphnew5,graphnew6,graphnew7}. Graphene is a one-atom thick layer of carbon atoms equipped with a hexagonal lattice structure where electrons obey a Dirac equation and have a linear dispersion relation with Fermi velocity $v_F$ \cite{graphnew1}. The hexagonal lattice brings along a number of topologically and geometrically originating new phenomena \cite{review} that rendered graphene an experimental material where ideas coming from 2+1 gravity can be tested \cite{graphcurv}. With respect to the latter perspective, it is the existence of multilayers and their intersection that create singularities, where physical quantities blow up. At these singular points, curvature is singular and the surfaces can be modelled by using a metric \cite{graphcurv}, so contact with $2+1$ gravity can be achieved. Graphene was identified for the first time in the laboratory in 2004 \cite{graphexper} and since then has created a new research stream for many theoretical ideas. For a recent review see \cite{review} and references therein. 

One of the most interesting phenomena in graphene is the localization of Dirac electrons and all the related problems on how to achieve localization \cite{graphnew1,wilzek,graphnew6,graphnew7}. Localization of fermions always occurs in the presence of defects and the first study was done in the seminal paper of Jackiw and Rossi \cite{graphzero}. The topology of the physical system's configuration space is critically affected by the presence of the defect, and this plays an important role for the localization effect. The localized fermionic modes can be classified according to a topological index theorem \cite{graphindex} which relates the net number of fermionic modes with the topology of the state space. In this article we shall be interested in graphene layers that have two types of defects, namely gapped graphene in the presence of a domain wall \cite{graphnew1} and superconducting graphene \cite{wilzek} in the presence of multivortices. In both frameworks there exist localized fermionic modes near the defects. Our study will be focused on another field theoretic aspect of graphene localized fermions, which is the existence of a rich extended supersymmetric structure underlying the system of localized fermions. As we shall explicitly demonstrate, these one dimensional supersymmetries have non-trivial topological charges, a fact probably indicating the existence of a non-linear supersymmetry. The relation of supersymmetry and graphene was also pointed out in \cite{graphenesusy}, but from a completely different point of view. In our study the focus will be on revealing the supersymmetric structures in both gapped and superconducting graphene and relating the corresponding Witten index with the localized modes. As we shall see, the lowest supersymmetric structure that underlies both systems is a number of distinct unbroken $N=2$, $d=1$ supersymmetries. The fact that the supersymmetries are unbroken is closely related to the existence of localized modes on the defects, but the proof for this is different in the two systems under study. The $N=2$ supersymmetries are combined to form $N$-extended supersymmetries with non-trivial topological charges. These extended supersymmetries are not simply higher order representations but are actually new supersymmetric realizations of the two systems. We shall see that supersymmetries remain unbroken even if the systems are perturbed by compact odd perturbations. This can help us to further understand the topological properties of the two systems, since from a physical point of view, compact perturbations can be caused by changing the pairing gap function $\Delta(r)$, with this perturbation leaving the Witten index of the system invariant. We believe that our study could provide another important field theoretic aspect of graphene-defects systems.

This paper is organized as follows: In section 1 we present the $N=2$ supersymmetric structure of the gapped graphene system along with some implications to the Hilbert space of the localized electrons. In section 2 we present the $N=4$ extended supersymmetric structure of the gapped graphene localized electrons and we also give a brief account on the non-reducible representations of $N=4$, $d=1$ supersymmetry. In section 3, we study what is the effect of domain wall perturbations on the Witten index and also if the Witten index changes when the gap function is changed. In section 4 we study the underlying supersymmetric structure in a superconducting graphene framework. The conclusions follow in the end of the paper.

\section{Gapped Graphene and One Dimensional Extended Supersymmetry-Non-Fredholm Operators Case}

\subsection{A Brief Gapped Graphene Primer}

A useful modification in graphene constructions, is to introduce an energy gap between the electron energies \cite{graphnew1}. This for example can be realized by using a staggered chemical potential \cite{graphnew1}. As it was shown in \cite{graphnew1}, domain walls can be materialized in a realistic way in graphene and these give rise to a band of mid-gap electron states. Specifically, the domain walls practically imitate an one dimensional metal embedded in a semi-conductor and thereby can be used as a single-channel quantum wire. The mid-gap states contain localized fermion modes in the Dirac Hamiltonian spectrum of the quantum system. These localized fermionic modes exist in the presence of topological defects like domains walls. The focus in this article with regards to gapped graphene, is on domain walls, which was studied in detail in \cite{graphnew1}. We shall adopt the notation of \cite{graphnew1}, in our study of the gapped graphene fermionic system. In \cite{graphnew1} it was shown that, owing to the existence of the domain wall, localized fermions occur in the  location of the domain wall. As we shall demonstrate in this section, there exists a rich one dimensional supersymmetric structure underlying the system of localized mid-gap electron states. 

The two valley electrons of graphene are described by the following Hamiltonian \cite{graphnew1},
\begin{equation}\label{hamgraph}
H_{graph}=\hbar v_F\left ( \begin{array}{ccccc}
  \frac{m(x)v_F}{\hbar} & i\frac{\mathrm{d}}{\mathrm{d}x}+ \frac{\mathrm{d}}{\mathrm{d}y}& 0 & 0\\
  i\frac{\mathrm{d}}{\mathrm{d}x}- \frac{\mathrm{d}}{\mathrm{d}y} & -\frac{m(x)v_F}{\hbar} & 0 & 0 \\
  0 & 0 & \frac{m(x)v_F}{\hbar} & i\frac{\mathrm{d}}{\mathrm{d}x}- \frac{\mathrm{d}}{\mathrm{d}y} \\
  0 & 0 & i\frac{\mathrm{d}}{\mathrm{d}x}+ \frac{\mathrm{d}}{\mathrm{d}y} & \frac{m(x)v_F}{\hbar} \\
\end{array}\right )
\end{equation}
The two graphene valleys are described by the diagonal blocks, which can be transformed to each other by time reversal and parity transformation. The domain wall is described by the function $m(x)$ which has a solitonic profile of the following form,
\begin{equation}\label{solitonprofilemass}
\lim_{x\rightarrow -\infty}m(x)=-m<0,{\,}{\,}{\,}\lim_{x\rightarrow \infty}m(x)=m>0
\end{equation}
We shall make a simple assumption that there exists at least one mid-gap electron state for each graphene valley, without loss of generality. This is enough to reveal the underlying supersymmetric structure.

\subsection{$N=2$, $d=1$ Supersymmetric Subalgebras}

The supersymmetries we shall present are one dimensional supersymmetries, also known as the research field of supersymmetric quantum mechanics. The latter \cite{reviewsusyqm} was introduced as a simplified model for the study of supersymmetry breaking in quantum field theory and nowadays is an independent research field, with many applications in various research areas. For example, in \cite{diffgeomsusyduyalities} and \cite{susyqminquantumsystems} interesting Hilbert space properties of supersymmetric quantum mechanical systems were studied, along with applications and also non linear realizations of supersymmetry. Some applications of supersymmetry in scattering related phenomena were studied in \cite{susyqmscatter} and various features of supersymmetry breaking were presented in \cite{susybreaking}. In addition, one dimensional supersymmetries are of great importance since, higher $N$-extended one dimensional supersymmetries \cite{extendedsusy} have a link to harmonic superspaces, see for example \cite{ivanov}. For some important works on supersymmetric quantum field theory see \cite{witten1,odi1,odi2,odi3} and references therein.

Before we reveal the extended one dimensional supersymmetric algebra that underlies the fermionic system on the gapped graphene domain wall, it is of crucial importance to present the four unbroken one dimensional $N=2$ subalgebras that underlie the system. It's importance is owing to the fact that these four algebras actually combine to form a higher non-trivial supersymmetric algebra and note that we are not discussing simply the formation of a higher reducible representation of the two simple algebras. The latter is ensured, as we explicitly demonstrate, by the existence of non-trivial topological supercharges.

To start with, consider the Hamiltonian (\ref{hamgraph}) in the limits described in relation (\ref{solitonprofilemass}). The corresponding Dirac equation reads,
\begin{equation}\label{shroe}
H_{graph}\psi= E\psi
\end{equation}
which can in turn be written in terms of two operator equations, in terms of the operators $\mathcal{D}_1$ and $\mathcal{D}_2$, given by,
\begin{equation}\label{susyqmrn5safsfsf67m}
\mathcal{D}_{1}=\left(%
\begin{array}{cc}
\frac{(m-E)v_F}{\hbar} & i\frac{\mathrm{d}}{\mathrm{d}x}+ \frac{\mathrm{d}}{\mathrm{d}y}
 \\  i\frac{\mathrm{d}}{\mathrm{d}x}- \frac{\mathrm{d}}{\mathrm{d}y} & -\frac{(m+E)v_F}{\hbar} \\
\end{array}%
\right),{\,}{\,}{\,}\mathcal{D}_{2}=\left(%
\begin{array}{cc}
\frac{(m-E)v_F}{\hbar} & i\frac{\mathrm{d}}{\mathrm{d}x}- \frac{\mathrm{d}}{\mathrm{d}y}
 \\  i\frac{\mathrm{d}}{\mathrm{d}x}+ \frac{\mathrm{d}}{\mathrm{d}y} & -\frac{(m+E)v_F}{\hbar} \\
\end{array}%
\right)
\end{equation}
 which are considered to act in the following two 2-component spinors $\psi_1$ and $\psi_2$ as follows,
\begin{equation}\label{twocompbispinors}
\mathcal{D}_{1}\psi_1=\mathcal{D}_{1}\left(%
\begin{array}{c}
  u_{-} \\
  u_{+} \\
\end{array}%
\right)=0,{\,}{\,}{\,}\mathcal{D}_{2}\psi_2=\mathcal{D}_{2}\left(%
\begin{array}{c}
  v_{-} \\
  v_{+} \\
\end{array}%
\right)=0
\end{equation}
The spinor $\psi$ is written in term of the 2-component spinors $\psi_1$ and $\psi_2$ in the following way,
\begin{equation}\label{bispinor}
\psi=\left(%
\begin{array}{c}
  \psi_1 \\
  \psi_2 \\
\end{array}%
\right)
\end{equation}
It is worth writing down the zero modes equations for the adjoint operators $\mathcal{D}_{1,2}$, which will prove to be useful later on in this section. The operators $\mathcal{D}_1^{\dag}$ and $\mathcal{D}_2^{\dag}$, satisfy the following equations,
\begin{equation}\label{diffeqnforadj1}
\mathcal{D}_1^{\dag}\psi_3=0,{\,}{\,}{\,}\mathcal{D}_2^{\dag}\psi_4=0
\end{equation}
By looking the form of the operators $\mathcal{D}_1$ and $\mathcal{D}_2$ in relation (\ref{susyqmrn5safsfsf67m}), we can easily verify that the exact form of the vectors $\psi_3$ and $\psi_4$ is,
\begin{equation}\label{formpsi34}
\psi_3=\left(%
\begin{array}{c}
  -v_{-} \\
  v_{+} \\
\end{array}%
\right),{\,}{\,}{\,}\psi_4=\left(%
\begin{array}{c}
  -u_{-} \\
  u_{+} \\
\end{array}%
\right)
\end{equation}
So practically speaking, the zero modes of the operators $\mathcal{D}_1^{\dag}$ and $\mathcal{D}_2^{\dag}$ are exactly the same as these of the operators $\mathcal{D}_1$ and $\mathcal{D}_2$, and particularly,
\begin{equation}\label{dimeke1r11firstrelation}
\mathrm{dim}{\,}\mathrm{ker}\mathcal{D}_{1}^{\dag}=\mathrm{dim}{\,}\mathrm{ker}\mathcal{D}_{2},{\,}{\,}{\,}\mathrm{dim}{\,}\mathrm{ker}\mathcal{D}_{2}^{\dag}=\mathrm{dim}{\,}\mathrm{ker}\mathcal{D}_{1}
\end{equation}
Relation (\ref{dimeke1r11firstrelation}) shall be proven quite useful when we shall address the issue of whether supersymmetry is unbroken, later on in this section.

Notice that from the previous section, the energy eigenvalue $E$ takes two values, namely,
\begin{equation}\label{energy}
E=\pm v_F\sqrt{k^2+m^2v_F^2}
\end{equation} 
so everything that follows is assumed to hold true for each existing energy eigenvalue. Let us focus to the operator $\mathcal{D}_1$ first and we demonstrate that we can built an $N=2$, $d=1$ algebra with it's basic constituent being the operator $\mathcal{D}_1$. Indeed, the supercharge $\mathcal{Q}_1$ and $\mathcal{H}_1$ of the $N=2$, $d=1$ superalgebra written in terms of the operator $\mathcal{D}_1$,
\begin{equation}\label{s7gsgdsgdgrdd}
\mathcal{Q}_{1}=\bigg{(}\begin{array}{ccc}
  0 & \mathcal{D}_{1} \\
  0 & 0  \\
\end{array}\bigg{)},{\,}{\,}{\,}\mathcal{Q}^{\dag}_{1}=\bigg{(}\begin{array}{ccc}
  0 & 0 \\
  \mathcal{D}_{1}^{\dag} & 0  \\
\end{array}\bigg{)},{\,}{\,}{\,}\mathcal{H}_{1}=\bigg{(}\begin{array}{ccc}
 \mathcal{D}_{1}\mathcal{D}_{1}^{\dag} & 0 \\
  0 & \mathcal{D}_{1}^{\dag}\mathcal{D}_{1}  \\
\end{array}\bigg{)}
\end{equation}
which satisfy the relations,
\begin{equation}\label{relationsforsusysddssdg}
\{\mathcal{Q}_{1},\mathcal{Q}^{\dag}_{1}\}=\mathcal{H}_{1}{\,}{\,},\mathcal{Q}_{1}^2=0,{\,}{\,}{\mathcal{Q}_{1}^{\dag}}^2=0
\end{equation}
These relations (\ref{relationsforsusysddssdg}) are the constituting equations of an $N=2$, $d=1$ algebra \cite{reviewsusyqm}. The issue whether supersymmetry is broken or not is a bit more involved and shall be properly addressed later on in this section and as we explicitly demonstrate, supersymmetry is unbroken, so we take that for granted for the moment. It is worth presenting what are the implications of unbroken supersymmetry for the Hilbert space that consists of the gapped graphene fermion states. We denote the Hilbert space of the supersymmetric quantum mechanical system as $\mathcal{H}_{sp}$, which is rendered a $Z_2$ graded space, by the action of the involution operator $\mathcal{W}$. This operator is called Witten parity, and it satisfies,
\begin{equation}\label{s45}
[\mathcal{W},\mathcal{H}_{sp}]=0,{\,}{\,}{\,}\{\mathcal{W},\mathcal{Q}_{1}\}=\{\mathcal{W},\mathcal{Q}_{1}^{\dag}\}=0
\end{equation}
Moreover, the Witten parity satisfies the following identity,
\begin{equation}\label{s6}
\mathcal{W}^{2}=1
\end{equation}
which is a property very common to projective operators. The operator $\mathcal{W}$ has the following matrix form representation in the case at hand,
\begin{equation}\label{wittndrf}
\mathcal{W}=\bigg{(}\begin{array}{ccc}
  1 & 0 \\
  0 & -1  \\
\end{array}\bigg{)}
\end{equation}
The actual operation of the Witten parity on the Hilbert space of the supersymmetric quantum system is that it spans the total Hilbert space into two $Z_2$ equivalent subspaces. As a consequence, the total Hilbert space of the quantum system acquires the following decomposition \cite{reviewsusyqm},
\begin{equation}\label{fgjhil}
\mathcal{H}_{sp}=\mathcal{H}^+\oplus \mathcal{H}^-
\end{equation}
with the vector-states belonging to the subspaces $\mathcal{H}^{\pm}$ being classified to even and odd parity states, according to their Witten parity, that is:
\begin{equation}\label{shoes}
\mathcal{H}^{\pm}=\mathcal{P}^{\pm}\mathcal{H}_{sp}=\{|\psi\rangle :
\mathcal{W}|\psi\rangle=\pm |\psi\rangle \}
\end{equation}
The decomposition of the Hilbert space has a direct implication on the total Hamiltonian $\mathcal{H}_1$, which is written as follows,
\begin{equation}\label{h1}
{\mathcal{H}}_{+}=\mathcal{D}_{1}{\,}\mathcal{D}_{1}^{\dag},{\,}{\,}{\,}{\,}{\,}{\,}{\,}{\mathcal{H}}_{-}=\mathcal{D}_{1}^{\dag}{\,}\mathcal{D}_{1}
\end{equation}
The operator $\mathcal{P}$, introduced in (\ref{shoes}) actually makes the classification of the Hilbert vectors to odd states and even states, since it's eigenstates $|\psi^{\pm}\rangle$, satisfy the following relation,
\begin{equation}\label{fd1}
P^{\pm}|\psi^{\pm}\rangle =\pm |\psi^{\pm}\rangle
\end{equation}
We shall call them for brevity positive and negative parity eigenstates \cite{reviewsusyqm}, with the term ''parity'' referring to the operator $P^{\pm}$. Using the Witten parity operator in the representation (\ref{wittndrf}), the parity eigenstates can acquire the following vector representation,
\begin{equation}\label{phi5}
|\psi^{+}\rangle =\left(%
\begin{array}{c}
  |\phi^{+}\rangle \\
  0 \\
\end{array}%
\right),{\,}{\,}{\,}
|\psi^{-}\rangle =\left(%
\begin{array}{c}
  0 \\
  |\phi^{-}\rangle \\
\end{array}%
\right)
\end{equation}
with $|\phi^{\pm}\rangle$ $\epsilon$ $\mathcal{H}^{\pm}$. We can write the vectors we defined in the relations above, in terms of the spinors $\psi_1$ and $\psi_2$. Indeed, the reader can convince himself that we can write,
\begin{equation}\label{fdgdfgh}
\psi_1 =|\phi^{-}\rangle=\left(%
\begin{array}{c}
  u_{-} \\
  u_{+} \\
\end{array}%
\right),{\,}{\,}{\,}\psi_3 =|\phi^{+}\rangle=\left(%
\begin{array}{c}
  -v_{-} \\
  v_{+} \\
\end{array}%
\right)
\end{equation}
Hence, we can write the corresponding even and odd parity supersymmetric quantum states in term of the vectors $\psi_1$ and $\psi_3$ as follows,
\begin{equation}\label{phi5}
|\psi^{+}\rangle =\left(%
\begin{array}{c}
  \psi_3 \\
  0 \\
\end{array}%
\right),{\,}{\,}{\,}
|\psi^{-}\rangle =\left(%
\begin{array}{c}
  0 \\
  \psi_1 \\
\end{array}%
\right)
\end{equation}
on which, the Hamiltonian and the supercharges of the supersymmetric algebra act. 

Having established the fact that a supersymmetric algebra can be constructed using operator $\mathcal{D}_1$, we can easily show by using the same line of argument, that another $N=2$, $d=1$ supersymmetric algebra can be constructed using the operator $\mathcal{D}_2$. Indeed, the supercharges and the Hamiltonian of this algebra are,
\begin{equation}\label{s7gsgdsgdgrddfffg}
\mathcal{Q}_{2}=\bigg{(}\begin{array}{ccc}
  0 & \mathcal{D}_{2} \\
  0 & 0  \\
\end{array}\bigg{)},{\,}{\,}{\,}\mathcal{Q}^{\dag}_{2}=\bigg{(}\begin{array}{ccc}
  0 & 0 \\
  \mathcal{D}_{2}^{\dag} & 0  \\
\end{array}\bigg{)},{\,}{\,}{\,}\mathcal{H}_{2}=\bigg{(}\begin{array}{ccc}
 \mathcal{D}_{2}\mathcal{D}_{2}^{\dag} & 0 \\
  0 & \mathcal{D}_{2}^{\dag}\mathcal{D}_{1}  \\
\end{array}\bigg{)}
\end{equation}
which satisfy the $N=2$, $d=1$ supersymmetric algebra:
\begin{equation}\label{relationsforsusysddghhfdssdgfffggf}
\{\mathcal{Q}_{2},\mathcal{Q}^{\dag}_{2}\}=\mathcal{H}_{2}{\,}{\,},\mathcal{Q}_{2}^2=0,{\,}{\,}{\mathcal{Q}_{2}^{\dag}}^2=0
\end{equation}
The rest of the analysis is the same and we omit it, for the shake of brevity. 

Now comes the issue of whether this supersymmetry is unbroken, which is very closely related to the following observation: The zero modes of the operator $\mathcal{D}_1$ are exactly the eigenfunctions of the gapped graphene fermionic system. Notice that this holds true for all energy eigenvalues, taken into account one at a time (the operator $\mathcal{D}_1$ is built using only one of them at a time, namely $E$). Of course the same applies for the operator $\mathcal{D}_2$. As we shall see, the zero modes of the operators $\mathcal{D}_1$ and $\mathcal{D}_2$ play a crucial role in the determination of whether supersymmetry is unbroken or not, due to the existence of an index theorem. Recall that supersymmetry is unbroken if the Witten index is a non-zero integer. The Witten index for Fredholm operators is equal to,
\begin{equation}\label{phil}
\Delta =n_{-}-n_{+}
\end{equation}
with $n_{\pm}$ the exact number of zero
modes of the operators ${\mathcal{H}}_{\pm}$ in the subspace $\mathcal{H}^{\pm}$. Notice that when Fredholm operators are considered, the zero modes have to be finitely many. In the case at hand, the operators are not Fredholm since the energy parameter takes values that span a continuum range, hence we have to make use of a continuum generalized Witten index. In a later section, when we study the superconducting graphene case, we shall come back to the issue of Fredholm operators and supersymmetry breaking. Let us focus on the first $N=2$ and the operator $\mathcal{D}_1$, for which the heat-kernel regularized index, denoted as $\mathrm{ind}_t\mathcal{D}_1$ and the Witten index, denoted as $\Delta_t$, are formally defined as follows \cite{reviewsusyqm,thaller},
\begin{align}\label{heatkerw}
& \mathrm{ind}_t\mathcal{D}_1=\mathrm{Tr}(-\mathcal{W}e^{-t\mathcal{D}_1^{\dag}\mathcal{D}_1})=\mathrm{tr}_{-}(-\mathcal{W}e^{-t\mathcal{D}_1^{\dag}\mathcal{D}_1})-\mathrm{tr}_{+}(-\mathcal{W}e^{-t\mathcal{D}_1\mathcal{D}_1^{\dag}}) \\ \notag 
& \Delta_t=\lim_{t\rightarrow \infty}\mathrm{ind}_t\mathcal{D}_1.
\end{align}
where we assumed that $t>0$, and also that $\mathrm{tr}_{\pm }$ denotes the trace in the subspaces $\mathcal{H}^{\pm}$. The formal definition of the heat-kernel regularized index involves trace-class operators, which have a finite norm \cite{thaller}. In the case at hand, the operator that must be trace-class is $\mathrm{tr}(-\mathcal{W}e^{-t\mathcal{D}_1^{\dag}\mathcal{D}_1})$. Now recall relation (\ref{dimeke1r11firstrelation}), from which we can easily establish the result that,
\begin{equation}\label{keeerrrr}
\mathrm{ker}\mathcal{D}_1=\mathrm{ker}\mathcal{D}_1^{\dag}\neq 0,
\end{equation}
owing to the existence of gapped graphene fermionic states. Relation (\ref{keeerrrr}) implies that, 
\begin{equation}\label{keeerrrr1}
\mathrm{ker}\mathcal{D}_1\mathcal{D}_1^{\dag}=\mathrm{ker}\mathcal{D}_1\mathcal{D}_1^{\dag}\neq 0,
\end{equation}
and consequently, the following relation holds true, regarding the operators
$e^{-t\mathcal{D}_1^{\dag}\mathcal{D}_1}$ and $e^{-t\mathcal{D}_1\mathcal{D}_1^{\dag}}$, 
\begin{equation}\label{qggeeee123}
\mathrm{tr}_{-}e^{-t\mathcal{D}_1^{\dag}\mathcal{D}_1}=\mathrm{tr}_{+}e^{-t\mathcal{D}_1\mathcal{D}_1^{\dag}}
\end{equation}
Recalling that $\mathrm{tr}_{\pm }$ denotes the trace corresponding to the subspaces $\mathcal{H}^{\pm}$,  relation (\ref{qggeeee123}) implies that the regularized index of the operator $\mathcal{D}_{1}$ is actually equal to zero. As a consequence, the regularized Witten index is also zero and in conjunction with relation (\ref{keeerrrr}) we come to the conclusion that the $N=2$ supersymmetric algebra corresponding to the operator $\mathcal{D}_1$ is unbroken. The same argument holds true for the algebra built on the operator $\mathcal{D}_2$, hence finally we have two unbroken $N=2$, $d=1$ supersymmetric algebras corresponding to the energy eigenvalue $E$. Bearing in mind that there are another two corresponding to the energy eigenvalue $-E$, we end up having four unbroken $N=2$, $d=1$ supersymmetries. Now the question is whether these supersymmetries combine in some way to form a higher order extended non-trivial supersymmetry. This is the subject of the next section.

\subsection{Global R-Symmetries}

The $N=2$ supersymmetric quantum mechanics algebra has some implications for the Hilbert space of the fermionic states that are localized on the domain wall. Specifically, it implies a global R-symmetry as we explicitly demonstrate. Focusing on the first algebra, described by the supercharges $\mathcal{Q}_1$ and $\mathcal{Q}_2$, the supercharges algebra is invariant under the following transformations:
\begin{align}\label{transformationu1}
& \mathcal{Q}_{1}=e^{-ia}\mathcal{Q'}_{1}, {\,}{\,}{\,}{\,}{\,}{\,}{\,}{\,}
{\,}{\,}\mathcal{Q}_{1}^{\dag}=e^{ia}\mathcal{Q'}_{1}^{\dag}
\end{align}
Therefore, the $N=2$ supersymmetric system is globally invariant under the global-$U(1)$ symmetry of relation (\ref{transformationu1}). This global
$U(1)$ symmetry is inherited to the Hilbert states
corresponding to the spaces $\mathcal{H}^{+}$,
$\mathcal{H}^{-}$ we presented earlier. Let, $\psi^{+}_{1}$ and
$\psi^{-}_{1}$ represent the Hilbert states corresponding to the graded Hilbert
spaces $\mathcal{H}^{+}$ and $\mathcal{H}^{-}$. Then, the $U(1)$ transformation transforms the Hilbert space states as follows,
\begin{equation}
\psi^{'+}_{1}=e^{-i\beta_{+}}\psi^{+}_{1},
{\,}{\,}{\,}{\,}{\,}{\,}{\,}{\,}
{\,}{\,}\psi^{'-}_{1}=e^{-i\beta_{-}}\psi^{-}_{1}
\end{equation}
where $\beta_{+}$ and $\beta_{-}$ are global parameters satisfying $a=\beta_{+}-\beta_{-}$. Having in mind that there are two $N=2$ supersymmetries for the energy eigenvalue $E$ and two for $-E$, the total $R$ symmetry, denoted as $R_{tot}$, of the localized fermionic system, is a product of four distinct $U(1)$ symmetries, that is,
\begin{equation}\label{fouru1s}
R_{tot}=U(1)\times U(1)\times U(1)\times U(1)
\end{equation}

\section{Topological Charge Extended $N=4$ Superalgebras in Gapped Graphene}

As we explicitly demonstrated, the domain wall graphene fermionic modes with energy $E$ constitute two separate $N=2$, $d=1$ algebras. We shall now demonstrate that these two supersymmetric algebras can be combined to give an $N=4$ extended supersymmetric algebra with non-trivial topological charges. This extended supersymmetric structure is not a simple higher dimensional representation of the supersymmetry algebra but a non-trivial supersymmetry with higher $N$ and non-trivial topological charges, which are not however central charges. In order to reveal the inherent extended supersymmetric structure of the system under study, we compute the following commutation and anticommutation relations that the supercharges and the Hamiltonians of the two algebras satisfy,
\begin{align}\label{commutatorsanticommedfgdfgrgdxvxtgrs}
&\{{{\mathcal{Q}}_{1}},{{\mathcal{Q}}^{\dag}_{1}}\}=2\mathcal{H}+\mathcal{Z}_{11},
{\,}\{{{\mathcal{Q}}_{2}},{{\mathcal{Q}}^{\dag}_{2}}\}=2\mathcal{H}+\mathcal{Z}_{22}
,{\,}\{{{\mathcal{Q}}_{2}},{{\mathcal{Q}}_{2}}\}=0,
\\ \notag &\{{{\mathcal{Q}}_{1}},{{\mathcal{Q}}_{ 1}}\}=0, {\,} \{{{\mathcal{Q}}_{2}},{{\mathcal{Q}}_{ 1}}^{\dag}\}=\mathcal{Z}_{2 1},{\,}\{{{\mathcal{Q}}_{1}},{{\mathcal{Q}}^{\dag}_{2}}\}=\mathcal{Z}_{1 2},{\,}\\ \notag
&\{{{\mathcal{Q}}^{\dag}_{1}},{{\mathcal{Q}}^{\dag}_{1}}\}=0,\{{{\mathcal{Q}}^{\dag}_{2}},{{\mathcal{Q}}^{\dag}_{2}}\}=0,\\
\notag
&[{{\mathcal{Q}}_{1}},{{\mathcal{Q}}_{2}}]=0,[{{\mathcal{Q}}^{\dag}_{2}},{{\mathcal{Q}}^{\dag}_{1}}]=0,{\,}[{{\mathcal{Q}}_{2}},{{\mathcal{Q}}_{1}}]=0,{\,}
\end{align}
These relations constitute an extended $N=4$, $d=1$ superalgebra with four non-trivial topological charges which we denoted as, $\mathcal{Z}_{11},\mathcal{Z}_{22},\mathcal{Z}_{12},\mathcal{Z}_{21}$ and with a Hamiltonian $\mathcal{H}$. The Hamiltonian is equal to,
\begin{equation}\label{hamgrapheaven}
\mathcal{H}=\mathrm{diag}\left (  \Delta_1 ;\Delta_2 ; \Delta_2 ;\Delta_1 \right )
\end{equation}
with $\Delta_1$ and $\Delta_2$ being equal to,
\begin{align}\label{diagmatxixhamilt}
& \Delta_1=\frac{\mathrm{d^2}}{\mathrm{d}x^2}+\frac{\mathrm{d^2}}{\mathrm{d}y^2}+\frac{(m-E)^2v_F^2}{\hbar^2} \\ \notag &
\Delta_2=\frac{\mathrm{d^2}}{\mathrm{d}x^2}+\frac{\mathrm{d^2}}{\mathrm{d}y^2}+\frac{(m+E)^2v_F^2}{\hbar^2}
\end{align}
The topological charge $\mathcal{Z}_{11}$ is equal to,
\begin{equation}\label{newsusymat22newsuperch}
\mathcal{Z}_{11 }=\left ( \begin{array}{ccccc}
  \mathcal{Z}^1_{11 } & 0 \\
  0 & \mathcal{Z}^2_{11 }  \\
\end{array}\right ).
\end{equation}
and the operator $\mathcal{Z}^1_{11 }$ stands for the following matrix,
\begin{equation}\label{newsusymat32sjhiuhguhg}
\mathcal{Z}^1_{11 }=\left ( \begin{array}{cc}
 i\frac{\mathrm{d^2}}{\mathrm{d}x\mathrm{d}y}-i\frac{\mathrm{d^2}}{\mathrm{d}y\mathrm{d}x} & (i\frac{\mathrm{d}}{\mathrm{d}x}+\frac{\mathrm{d}}{\mathrm{d}y})\frac{(m+E)v_F}{\hbar}+\frac{(m-E)v_F}{\hbar}(-i\frac{\mathrm{d}}{\mathrm{d}x}-\frac{\mathrm{d}}{\mathrm{d}y}) \\ -\frac{(m+E)v_F}{\hbar}(-i\frac{\mathrm{d}}{\mathrm{d}x}+\frac{\mathrm{d}}{\mathrm{d}y})+(i\frac{\mathrm{d}}{\mathrm{d}x}-\frac{\mathrm{d}}{\mathrm{d}y})\frac{(m-E)v_F}{\hbar} &  i\frac{\mathrm{d^2}}{\mathrm{d}y\mathrm{d}x}-i\frac{\mathrm{d^2}}{\mathrm{d}x\mathrm{d}y} \\
\end{array}\right )
\end{equation}
while the operator $\mathcal{Z}^2_{11 }$, is stands for,
\begin{equation}\label{newsusymat42sfgghuguohfdogudfg}
\mathcal{Z}^2_{11 }=\left ( \begin{array}{cc}
 -i\frac{\mathrm{d^2}}{\mathrm{d}x\mathrm{d}y}+i\frac{\mathrm{d^2}}{\mathrm{d}y\mathrm{d}x} & -(-i\frac{\mathrm{d}}{\mathrm{d}x}+\frac{\mathrm{d}}{\mathrm{d}y})\frac{(m+E)v_F}{\hbar}+\frac{(m-E)v_F}{\hbar}(-i\frac{\mathrm{d}}{\mathrm{d}x}-\frac{\mathrm{d}}{\mathrm{d}y}) \\ (i\frac{\mathrm{d}}{\mathrm{d}x}+\frac{\mathrm{d}}{\mathrm{d}y})\frac{(m+E)v_F}{\hbar}+\frac{(m-E)v_F}{\hbar}(i\frac{\mathrm{d}}{\mathrm{d}x}-\frac{\mathrm{d}}{\mathrm{d}y}) &  -i\frac{\mathrm{d^2}}{\mathrm{d}y\mathrm{d}x}+i\frac{\mathrm{d^2}}{\mathrm{d}x\mathrm{d}y} \\
\end{array}\right )
\end{equation}
Moreover, the topological charge $Z_{22}$ is,
\begin{equation}\label{newsusymat22newsuperch}
\mathcal{Z}_{11 }=\left ( \begin{array}{ccccc}
  \mathcal{Z}^1_{11 } & 0 \\
  0 & \mathcal{Z}^2_{11 }  \\
\end{array}\right ).
\end{equation}
with the operator $\mathcal{Z}^1_{22 }$ being equal to,
\begin{equation}\label{newsusymat32sjhiuhguhg}
\mathcal{Z}^1_{11 }=\left ( \begin{array}{cc}
 i\frac{\mathrm{d^2}}{\mathrm{d}x\mathrm{d}y}-i\frac{\mathrm{d^2}}{\mathrm{d}y\mathrm{d}x} & (i\frac{\mathrm{d}}{\mathrm{d}x}-\frac{\mathrm{d}}{\mathrm{d}y})\frac{(m-E)v_F}{\hbar}-\frac{(m+E)v_F}{\hbar}(-i\frac{\mathrm{d}}{\mathrm{d}x}+\frac{\mathrm{d}}{\mathrm{d}y}) \\ \frac{(m-E)v_F}{\hbar}(-i\frac{\mathrm{d}}{\mathrm{d}x}-\frac{\mathrm{d}}{\mathrm{d}y})-(i\frac{\mathrm{d}}{\mathrm{d}x}-\frac{\mathrm{d}}{\mathrm{d}y})\frac{(m+E)v_F}{\hbar} &  i\frac{\mathrm{d^2}}{\mathrm{d}y\mathrm{d}x}-i\frac{\mathrm{d^2}}{\mathrm{d}x\mathrm{d}y} \\
\end{array}\right )
\end{equation}
and in addition, the operator $\mathcal{Z}^2_{22 }$, is equal to,
\begin{equation}\label{newsusymat42sfgghuguohfdogudfg}
\mathcal{Z}^2_{11 }=\left ( \begin{array}{cc}
 -i\frac{\mathrm{d^2}}{\mathrm{d}x\mathrm{d}y}+i\frac{\mathrm{d^2}}{\mathrm{d}y\mathrm{d}x} & (i\frac{\mathrm{d}}{\mathrm{d}x}+\frac{\mathrm{d}}{\mathrm{d}y})\frac{(m-E)v_F}{\hbar}-\frac{(m-E)v_F}{\hbar}(i\frac{\mathrm{d}}{\mathrm{d}x}+\frac{\mathrm{d}}{\mathrm{d}y}) \\ -(i\frac{\mathrm{d}}{\mathrm{d}x}-\frac{\mathrm{d}}{\mathrm{d}y})\frac{(m+E)v_F}{\hbar}+\frac{(m-E)v_F}{\hbar}(i\frac{\mathrm{d}}{\mathrm{d}x}-\frac{\mathrm{d}}{\mathrm{d}y}) &  -i\frac{\mathrm{d^2}}{\mathrm{d}y\mathrm{d}x}+i\frac{\mathrm{d^2}}{\mathrm{d}x\mathrm{d}y} \\
\end{array}\right )
\end{equation}
Finally, the topological charge $\mathcal{Z}_{12 }$ is, 
\begin{equation}\label{newsusymat22newsuperchwt123}
\mathcal{Z}_{12 }=\left ( \begin{array}{ccccc}
  \mathcal{Z}^1_{12 } & 0 \\
  0 & \mathcal{Z}^2_{12 }  \\
\end{array}\right ).
\end{equation}
with $\mathcal{Z}^1_{12 }$ being equal to,
\begin{equation}\label{newsusymat32sjhiuhguhgwt}
\mathcal{Z}^1_{12 }=\left ( \begin{array}{cc}
 -\frac{\mathrm{d^2}}{\mathrm{d}x^2}-i\frac{\mathrm{d^2}}{\mathrm{d}x\mathrm{d}y}+i\frac{\mathrm{d^2}}{\mathrm{d}y\mathrm{d}x}-\frac{\mathrm{d^2}}{\mathrm{d}y^2}+\frac{(m-E)^2v_F}{\hbar^2} & -(i\frac{\mathrm{d}}{\mathrm{d}x}+\frac{\mathrm{d}}{\mathrm{d}y})\frac{(m+E)v_F}{\hbar}+\frac{(m-E)v_F}{\hbar}(i\frac{\mathrm{d}}{\mathrm{d}x}+\frac{\mathrm{d}}{\mathrm{d}y}) \\ -\frac{(m+E)v_F}{\hbar}(-i\frac{\mathrm{d}}{\mathrm{d}x}-\frac{\mathrm{d}}{\mathrm{d}y})+(i\frac{\mathrm{d}}{\mathrm{d}x}-\frac{\mathrm{d}}{\mathrm{d}y})\frac{(m-E)v_F}{\hbar} &  -\frac{\mathrm{d^2}}{\mathrm{d}x^2}+i\frac{\mathrm{d^2}}{\mathrm{d}x\mathrm{d}y}-i\frac{\mathrm{d^2}}{\mathrm{d}y\mathrm{d}x}-\frac{\mathrm{d^2}}{\mathrm{d}y^2}+\frac{(m+E)^2v_F}{\hbar^2} \\
\end{array}\right )
\end{equation}
and also, the operator $\mathcal{Z}^2_{12 }$, is equal to:
\begin{equation}\label{newsusymat42sfgghuguohfdogudfgwt1}
\mathcal{Z}^2_{11 }=\left ( \begin{array}{cc}
 -\frac{\mathrm{d^2}}{\mathrm{d}x^2}-i\frac{\mathrm{d^2}}{\mathrm{d}x\mathrm{d}y}-i\frac{\mathrm{d^2}}{\mathrm{d}y\mathrm{d}x}-\frac{\mathrm{d^2}}{\mathrm{d}y^2}+\frac{(m+E)^2v_F}{\hbar^2}  & -(i\frac{\mathrm{d}}{\mathrm{d}x}+\frac{\mathrm{d}}{\mathrm{d}y})\frac{(m-E)v_F}{\hbar}-\frac{(m+E)v_F}{\hbar}(i\frac{\mathrm{d}}{\mathrm{d}x}-\frac{\mathrm{d}}{\mathrm{d}y}) \\ -(-i\frac{\mathrm{d}}{\mathrm{d}x}+\frac{\mathrm{d}}{\mathrm{d}y})\frac{(m+E)v_F}{\hbar}+\frac{(m-E)v_F}{\hbar}(i\frac{\mathrm{d}}{\mathrm{d}x}+\frac{\mathrm{d}}{\mathrm{d}y}) &  \frac{\mathrm{d^2}}{\mathrm{d}x^2}+i\frac{\mathrm{d^2}}{\mathrm{d}x\mathrm{d}y}+i\frac{\mathrm{d^2}}{\mathrm{d}y\mathrm{d}x}-\frac{\mathrm{d^2}}{\mathrm{d}y^2}+\frac{(m-E)^2v_F}{\hbar^2} \\
\end{array}\right )
\end{equation}
The remaining topological charge $\mathcal{Z}_{21 }$ is the conjugate of $\mathcal{Z}_{12 }$, that is, $\mathcal{Z}_{12 }=\mathcal{Z}_{12 }^{\dag}$, so we omit the details.

Before we close this section a brief comment on the topological charges we found is in order. The appearance of non-trivial topological charges in supersymmetric algebras of any dimension was firstly noticed in \cite{wittentplc}, were actually the terminology topological charge was first used. Topological charges cannot be considered as central charges \cite{fayet}, since there exist non-vanishing commutation relations of these with some operators of the superalgebra. Intriguingly enough, the theoretical framework of reference \cite{wittentplc}, where topological charges where firstly pointed out, consisted of a supersymmetric algebra in the presence of topological defects. It seems that supersymmetry and topological charges have a deeper interconnection, as was also pointed out in \cite{topologicalcharges}, and also the existence of non-trivial topological charges in a supersymmetric algebra, could be the indicator of a non-linear and certainly non-trivial supersymmetric structure. We defer this investigation for a future work.

\subsection{Representations of the Algebras}

Having established the result that the gapped graphene fermions possess a rich supersymmetric structure which is at most an $N=4$ extended supersymmetry, it is normal to ask whether there can be any realistic structure formation at which this supersymmetry can actually be realized. In view of this question, we shall present in this section an indirect way of perhaps observing the supersymmetric structure. What we actually intend to do is to briefly present the irreducible representations of the supersymmetric algebra at hand. In this way, we do not actually find a realistic structure but since the gapped graphene fermions constitute this supersymmetry, perhaps the observation that these are classified according to a specific pattern, could be linked to our investigation. By irreducible representations, we do not mean the perspective we adopted in \cite{oikonomoudomain}, but we are interested in irreducible representations of $N$-extended supersymmetry.

It is very well known in the related to the subject literature \cite{toppannew}, that $N$-extended supersymmetry and the division algebras of real, complex quaternionic and octonionic numbers \cite{toppannew} are in close connection. The latter when applied to one dimensional supersymmetric algebras, then this close relation can be actually viewed as a correlation between Clifford algebras and the $N$-extended supersymmetric algebras. Clifford algebras have irreducible representations which are classified in terms of division algebras and more specifically octonions (see for example \cite{toppannew} for details). The classification of the N-extended one dimensional supersymmetry irreducible representations can be realized if the admissible ordered integers $(n_1,n_2,n_3,...,n_k)$ can be found for any given $N$. Then these admissible ordered integers correspond to the irreducible multiplets with $n_i$ fields of dimension $d_i$. Note that the length of the irreducible representation is the integer $m$ which corresponds to the maximum dimensionality of a representation $d_m$.

The complete and detailed study on the irreducible representations of extended supersymmetry was done in detail in references \cite{toppannew}, and specifically by Pashnev and Toppan (2001). We shall not go in details since the work is done in \cite{toppannew}, we just present the most sound results of these studies, corresponding of course to the $N=4$ case.

All the multiplets can be formed in such a way, so that these multiplets is in one to one correspondence with the set of reducible representations and in effect, each multiplet corresponds to only one representation. This can be achieved if an equivalence relation is used among the multiplets \cite{toppannew}. The representations contain in general, a number $n$ of bosonic and fermionic fields, with $n$ and $N$ being related as follows,
\begin{equation}\label{ehdhd}
N=8p+q,{\,}{\,}{\,}n=2^{4p}G(q)
\end{equation}
 with $p=0,1,2,...$ and $q=1,2,...,8$. The function $G(q)$, which is related to the 8 modulo Bott periodicity, is known as the Radon-Hurwitz function. Notice that the 8 modulo Bott periodicity is a consequence of the underlying octonionic structure. Given the number $N$ of extended supersymmetry, one can find length-3 representations which have the form $(n-k,n,k)$, with $k$ being a positive integer with values $k=0,1,2,..,2n$. In addition we can form length-4 representations, which however exist for specific extended supersymmetric structures, with the total number $N$ taking the values $N=3,5,7$ and also for $N\geq 9$. Higher length representations exist only for $N\geq 10$. In our case, we are interested in the case $N=4$ so we have at most length-3 representations and below we quote all the irreducible length-3
 representations,
\begin{equation}\label{4lengthirreps}
(4,4,0),{\,}{\,}{\,}(3,4,1),{\,}{\,}{\,}(2,4,2),{\,}{\,}{\,}(1,4,3)
\end{equation}
In addition to the above, it is possible to form tensor product representations using the representations (\ref{4lengthirreps}) but we refrain from going into further details, since these can found in \cite{toppannew}.

\section{Domain Wall Perturbations in Gapped Graphene and the Witten Index}

Having found a rich supersymmetric structure underlying the fermionic system corresponding to gapped graphene, in this section we shall study the effect of domain wall perturbations on the Witten index of the supersymmetric algebras. Recall that the domain wall effect was assumed to be a solitonic type effect described by relation (\ref{solitonprofilemass}). Now consider that we perturb this form by adding a slow varying function of $x$, so that the solitonic profile at infinity is nearly described by the following limits,
\begin{equation}\label{solitonprofilemass}
\lim_{x\rightarrow -\infty}m(x)=-m-m^{-\infty}(x)<0,{\,}{\,}{\,}\lim_{x\rightarrow \infty}m(x)=m+m^{\infty}(x)>0
\end{equation}
This behavior of the domain wall has a direct effect on the Dirac equation of the fermionic modes, which in turn has an impact on the operators $\mathcal{D}_1$ and $\mathcal{D}_2$ defined in (\ref{susyqmrn5safsfsf67m}). We focus for the moment on the former but the same argument applies for the latter too. The operator $\mathcal{D}_1$ is modified and we denote the new operator as $\mathcal{D}_{n}$, which has the following form,
\begin{equation}\label{susyqmrn5safsfsf67mnewrel}
\mathcal{D}_{n}=\left(%
\begin{array}{cc}
\frac{(m+m^{\infty}(x)-E)v_F}{\hbar} & i\frac{\mathrm{d}}{\mathrm{d}x}+ \frac{\mathrm{d}}{\mathrm{d}y}
 \\  i\frac{\mathrm{d}}{\mathrm{d}x}- \frac{\mathrm{d}}{\mathrm{d}y} & -\frac{(m^{-\infty}(x)+E)v_F}{\hbar} \\
\end{array}%
\right)
\end{equation}
which can be written in the following equivalent form,
\begin{equation}\label{bnewfoirf}
\mathcal{D}_{n}=\mathcal{D}_{1}+\mathcal{C}
\end{equation}
with $\mathcal{C}$,
\begin{equation}\label{codd}
\mathcal{C}=\left(%
\begin{array}{cc}
 0& \frac{m^{\infty}(x)v_F}{\hbar}
 \\ -\frac{m^{-\infty}(x)v_F}{\hbar} & 0\\
\end{array}%
\right)
\end{equation}
The operator $\mathcal{C}$ contains non-infinite terms, since the functions $m^{\pm \infty}(x)$ are slowly varying, and as a consequence of that, it is a bounded operator. In addition, it is an odd matrix and therefore the operators $\mathcal{D}_n$, $\mathcal{D}_1$ and $\mathcal{C}$, satisfy a theorem which states (see \cite{thaller} page 168, Theorem 5.28):

\begin{itemize}
 \item Let $D$ be a trace class operator and $C$ a bounded odd operator. Then, the regularized indices of $D+C$ and $C$ are equal, that is

\begin{equation}\label{indperturbhfgatrn}
\mathrm{ind}_{t}(D+C)=\mathrm{ind}_{t}D
\end{equation}
\end{itemize}
In the case at hand, the operator $\mathcal{D}_1$ is trace-class and as we saw, the operator $\mathcal{C}$ is bounded, so the theorem applies directly. As a consequence, we have that,
\begin{equation}\label{indperturbhhgjhjghkjgjfgatrn}
\mathrm{ind}_t\mathcal{D}_{n}=\mathrm{ind}_t(\mathcal{D}_{1}+\mathcal{C})=\mathrm{ind}_t\mathcal{D}_{1}
\end{equation}
Therefore, by recalling how the Witten index is connected to the regularized index of the operator $\mathcal{D}_1$ (see relation (\ref{heatkerw})), we conclude that the modification of the domain wall solitonic profile has no effect on the Witten index of the supersymmetric algebra that underlies the system, and thereby the $N=2$, $d=1$ supersymmetric algebra remains unbroken. The same applies of course for the rest three $N=2$ supersymmetric algebras.

We have to note that if we add Hopping effects \cite{graphnew1}, the Witten index will still remain invariant, but we omit this analysis since the line of the argument is pretty much the same as in the case we just presented.

\section{Superconducting Graphene and One Dimensional Extended Supersymmetries-Fredholm Operators Case}

It is well known that topological defects such as kinks, domain walls and vortices arise quite frequently in physical systems as excitations in a background quantum field theory or in an ordered state of matter \cite{graphnew6}. Fermions existing in these topological backgrounds can have fractional quantum numbers, with the fractionalization mediated by zero-energy bound states of the fermions to the defect. The number of zero modes is a central feature of these physical systems and helps towards the complete understanding of the physical properties of the physical system consisting of fermions and defects \cite{graphnew1,wilzek,graphnew6}. This number of zero modes is usually given in terms of an index theorem \cite{graphnew1,wilzek,graphnew6}.

In view of the importance of zero modes to defect-fermions physical systems, we shall study the zero modes of superconducting graphene and express the number of electron zero modes of graphene in terms of a supersymmetric Witten index. We will explicitly demonstrate that, as in the case of gapped graphene, in the superconducting graphene the electrons zero modes constitute unbroken $N=2$, $d=1$ algebras which actually combine to give an extended supersymmetric algebra, which is much more complicated in comparison to the gapped graphene. The difference between the two cases is that in the superconducting graphene case, the zero modes of the fermions constitute the supersymmetric algebras.

Superconductivity in graphene can be induced in multiple ways \cite{wilzek}, with the most simple way being the one involving multivortices \cite{wilzek}. The vortices in the superconducting graphene state acquire interesting internal structure \cite{wilzek}, because each vortex supports a low energy mode of the Bogoliubov-de-Gennes equations. These structured vortices resemble in some aspects vortices in superconductors, with the surface of topological insulators supporting Dirac zero modes. It is intriguing that a supersymmetric structure, quite similar but much more simple in comparison to the one we present here, exists in topological insulators, as was demonstrated in \cite{oikonomousuperconductors}.

The full details for the superconducting graphene can be found in the article by Ghaemi and Wilczek \cite{wilzek}, here we present a few necessary for our presentation information.  The electron zero modes in the presence of a vortex in superconducting graphene is described by the Bogoliubov-de-Gennes equation,
\begin{equation}\label{hamgraphsupercondgraphen}
\left ( \begin{array}{ccccc}
  H_+^p+H_+^A & 0 & \Delta (r) & 0\\
  0 & H_+^p+H_+^A & 0 & \Delta (r)  \\
  \Delta^* (r) & 0 & -H_+^p+H_+^A & i\frac{\mathrm{d}}{\mathrm{d}x}- \frac{\mathrm{d}}{\mathrm{d}y} \\
  0 & \Delta^* (r)  & 0 & -H_+^p+H_+^A \\
\end{array}\right )\Psi =0
\end{equation}
with $\Psi $ being equal to the 4-component spinor,
\begin{equation}\label{formpsi3434345}
\Psi=\left(%
\begin{array}{c}
  u_{-} \\
  u_{+} \\
  v_{-} \\
  v_{+} \\
\end{array}%
\right)
\end{equation}
and with the operators $H_{\pm}^p$ and $H_{\pm}^A$ being equal to,
\begin{equation}\label{oprs}
H_{\pm}^p=-i(\sigma_x\partial_x\pm\sigma_y\partial_y),{\,}{\,}{\,}H_{\pm}^A=-q(\sigma_xA_x\pm \sigma_yA_y)
\end{equation}
The $(A_x,A,y)$ denote the components of the electromagnetic vector potential which will describe the multivortices. Subscripts refer to valley index and pseudospin. The equations (\ref{hamgraphsupercondgraphen}) decouple in two independent sets involving $u_+,v_{-}$, with the two sets of solutions being related by a reflection about the $x$ axis, so we focus on the first set of equations. Note that the supersymmetries we will find are doubled in the end, due to the existence of these two sets of equations. By putting $v=\sigma_yu^*$, choosing $\vec{A}=-e_{\theta}A(r)$ and decomposing $u_{+}$ as follows,
\begin{equation}\label{formpsi3434345dec}
u_{+}=\left(%
\begin{array}{c}
  a(r) \\
  b(r) \\
\end{array}%
\right)
\end{equation}
we end up to the following set of equations,
\begin{align}\label{finaleqnsgraphe1}
& e^{i\theta}\left (\frac{\partial}{\partial r}+\frac{i}{r}\frac{\partial}{\partial \theta} a-q A(r)e^{i\theta}a+\Delta (r) a^*=0\right ) \\ \notag & -e^{i\theta}\left (\frac{\partial}{\partial r}-\frac{i}{r}\frac{\partial}{\partial \theta} b+q A(r)e^{i\theta}b+\Delta (r) b^*=0\right )
\end{align}
A rescaling can eliminate the vector field \cite{wilzek} and we assume that the condensate function $\Delta (r)$ has the form $\Delta (r)=\Delta_n(r)e^{in\theta}$. The function $\Delta_n$ is considered to behave as follows,
\begin{equation}\label{functionbehga}
\lim_{r\rightarrow 0}\Delta_n(r)\rightarrow r^{|n|},{\,}{\,}{\,}\lim_{r\rightarrow \infty}\Delta_n(r)\rightarrow \mathrm{const}
\end{equation}
a configuration that is appropriate to describe an $n$-fold multivortex \cite{wilzek}. We focus on the first equation of (\ref{finaleqnsgraphe1}), so the number of the final supersymmetries takes contribution from the $b(r)$ equation too. There are two kinds of solutions as was explicitly shown in \cite{wilzek} and we are interested in the one, in which $a(r)$ is decomposed as $a(r)=f(r)e^{il\theta}+g(r)e^{im\theta}$, with $l+m=n-1$. Assuming real functions we end up to the set of equations,
\begin{align}\label{seteqns}
& \frac{\mathrm{d}f(r)}{\mathrm{d}r}-\frac{l}{r}f(r)+\Delta_n(r)g(r)=0\\ \notag &
\frac{\mathrm{d}g(r)}{\mathrm{d}r}-\frac{m}{r}f(r)+\Delta_n(r)f(r)=0
\end{align}
Now the solutions are in order. For $n$ odd, there are $n-1$ zero modes for $a(r)$ with $n-1\geq l\geq 1$, while when $n$-even,  $n-1\geq l\geq 0$. As for $b(r)$, there are $n$ zero modes for $n\geq 0$ and none for $n<0$. All the zero modes are assumed to be normalizable.

\subsection{Construction of $N=2$, $d=1$ Supersymmetries}

Along the same line of research we followed in the gapped graphene case, we can easily construct the $N=2$, $d=1$ supersymmetries in a straightforward way. Our analysis shall be based on equation (\ref{seteqns}) which gives the zero modes of the superconducting graphene corresponding to the function $u_+$. For each set of $(n,m,l)$ we can construct an operator similar to $\mathcal{D}_1$ in the case of gapped graphene. We denote by $i$ the triad of numbers $i=(n,m,l)$ for simplicity, and the operator that can be constructed from (\ref{seteqns}) is the following,
\begin{equation}\label{dfsaskia}
\mathcal{D}_i=\left(%
\begin{array}{cc}
\frac{\mathrm{d}}{\mathrm{d}r}-\frac{l}{r} & \Delta_n(r)
 \\  \Delta_n(r) & \frac{\mathrm{d}}{\mathrm{d}r}-\frac{m}{r} \\
\end{array}%
\right)
\end{equation}
For each set of numbers $i$, we can form the following supercharges and Hamiltonians,
\begin{equation}\label{s7gsgdsgdgrddtriad}
\mathcal{Q}_{i}=\bigg{(}\begin{array}{ccc}
  0 & \mathcal{D}_{i} \\
  0 & 0  \\
\end{array}\bigg{)},{\,}{\,}{\,}\mathcal{Q}^{\dag}_{i}=\bigg{(}\begin{array}{ccc}
  0 & 0 \\
  \mathcal{D}_{i}^{\dag} & 0  \\
\end{array}\bigg{)},{\,}{\,}{\,}\mathcal{H}_{i}=\bigg{(}\begin{array}{ccc}
 \mathcal{D}_{i}\mathcal{D}_{i}^{\dag} & 0 \\
  0 & \mathcal{D}_{i}^{\dag}\mathcal{D}_{i}  \\
\end{array}\bigg{)}
\end{equation}
which satisfy the relations,
\begin{equation}\label{relationsforsusysddssdgtriad}
\{\mathcal{Q}_{i},\mathcal{Q}^{\dag}_{i}\}=\mathcal{H}_{i}{\,}{\,},\mathcal{Q}_{i}^2=0,{\,}{\,}{\mathcal{Q}_{i}^{\dag}}^2=0
\end{equation}
Therefore, since we have $n$ sets of such operators, we have $n$ distinct $N=2$, $d=1$ supersymmetries. Each one of these supersymmetries share all the characteristics we presented in the gapped graphene case, so we omit the details. In addition, each one of these supersymmetries is unbroken. This is easy to demonstrate and it is based on the fact that for each $n$ there are exactly $n-1$ zero modes. This means that the kernel of each of the operators is finite, and particularly,
\begin{equation}\label{kefrdi}
\mathrm{dim}{\,}\mathrm{ker}\mathcal{D}_i=n-1
\end{equation} 
In addition the kernel of the adjoint operator $\mathcal{D}_i^{\dag}$ consists of exactly the same zero modes as $\mathcal{D}_i$, since the latter operator is self adjoint, hence we have,
\begin{equation}\label{kefrdi1}
\mathrm{dim}{\,}\mathrm{ker}\mathcal{D}_i^{\dag}=n-1
\end{equation}
Since the operators $\mathcal{D}_i$ and $\mathcal{D}_i^{\dag}$ have a finite kernel, these are Fredholm operators. Now let us formally address the supersymmetry breaking issue for Fredholm operators. The supersymmetry breaking is controlled by the Witten index, which for the finite kernel case is defined as follows,
\begin{equation}\label{phil}
\Delta =n_{-}-n_{+}
\end{equation}
with $n_+$ and $n_{-}$ the number of zero modes of the operators $\mathcal{D}_i\mathcal{D}_i^{\dag}$ and $\mathcal{D}_i^{\dag}\mathcal{D}_i$ respectively. When the Witten index is a non-zero
integer, supersymmetry unbroken. The case when the Witten index is zero is a bit more complicated because when $n_{+}=n_{-}=0$
supersymmetry is obviously broken, but when $n_{+}= n_{-}\neq 0$
supersymmetry is not broken. The latter case applies for the operator $\mathcal{D}_i$. Indeed, recall that the Fredholm index of the operator $\mathcal{D}_i$ is equal to,
\begin{equation}\label{ker}
\mathrm{ind} \mathcal{D}_i = \mathrm{dim}{\,}\mathrm{ker}
\mathcal{D}_i-\mathrm{dim}{\,}\mathrm{ker} \mathcal{D}_i^{\dag}=
\mathrm{dim}{\,}\mathrm{ker}\mathcal{D}_i^{\dag}\mathcal{D}_i-\mathrm{dim}{\,}\mathrm{ker}\mathcal{D}_i\mathcal{D}_i^{\dag}
\end{equation}
Then, the Witten index is related to the Fredholm index as follows,
\begin{equation}\label{ker1}
\Delta=\mathrm{ind} \mathcal{D}_i=\mathrm{dim}{\,}\mathrm{ker}
\mathcal{D}_i^{\dag}\mathcal{D}_i-\mathrm{dim}{\,}\mathrm{ker} \mathcal{D}_i\mathcal{D}_i^{\dag}
\end{equation}
Owing to the fact that, 
\begin{equation}\label{kefrdishfdh}
\mathrm{dim}{\,}\mathrm{ker}\mathcal{D}_i=\mathrm{dim}{\,}\mathrm{ker}\mathcal{D}_i^{\dag}=n-1\neq 0
\end{equation} 
each of the $n$ different supersymmetries is unbroken. Bearing in mind that there is another set of solutions which we did not take into account (see below relation (\ref{oprs})), the final number of $N=2$, $d=1$ supersymmetries is actually $4n$. In the next section we focus on one type of these supersymmetries for simplicity and we shall reveal the extended supersymmetric structure of these $n$ $N=2$ supersymmetries.  

Before closing this section it is worth mentioning that owing to the fact that the operators we dealt with in this section are Fredholm, the compact perturbations of the operators leave the Witten index invariant. This is because the Fredholm operators are by definition trace-class. This means that if we perturb the gap function $\Delta(r)$ so that it produces compact perturbations, supersymmetry remains unbroken. A similar conclusion was derived in \cite{wilzek}, but in relation to the Fredholm index of the corresponding Dirac operators.

\subsection{Extended Supersymmetric Structure for $(u_+,v_{-})$ Subsystem}

As in the gapped graphene case, we shall investigate if the $n$ different supersymmetries we found in the previous section, combine to form an extended supersymmetric structure. As we shall explicitly demonstrate, these indeed combine to form a much more complicated structure, in comparison to the gapped graphene case, where we found only an $N=4$ supersymmetry with non trivial supercharges. In the present case that we study superconducting graphene, we can form a number of $n$ different supercharges of the following form:
\begin{equation}\label{wit2jdnhdgeneralc}
\mathcal{Q}_{i}=\bigg{(}\begin{array}{ccc}
  0 & \mathcal{D}_{i} \\
  0 & 0  \\
\end{array}\bigg{)}
\end{equation}
with $\mathcal{D}_{i}$ given in relation (\ref{dfsaskia}) and $i=1,2,3,...n$. The supercharges $\mathcal{Q}_i$ can form an $N=2n$ extended supersymmetric one dimensional algebra with non-trivial topological charges, which is described by the following algebra,
\begin{align}\label{n4algbe1sjdjfgeneraldfgdfg}
&\{Q_{i},Q_{j}^{\dag}\}=2\delta_{ij}\mathcal{H}+Z_{ij},{\,}{\,}{\,}{\,}i ,j=1,2,..n \\ \notag &
\{Q_{i},Q_{j}\}=0,{\,}{\,}\{Q_{i}^{\dag},Q_{j}^{\dag}\}=0
\end{align}
In the relation above, the generalized Hamiltonian $\mathcal{H}$ is equal to,
\begin{equation}\label{newsusymat121233}
\mathcal{H}=\left ( \begin{array}{ccccc}
  \frac{\mathrm{d}^2}{\mathrm{d}r^2} & 0 & 0 & 0\\
  0 & \frac{\mathrm{d}^2}{\mathrm{d}r^2} & 0 & 0 \\
  0 & 0 & \frac{\mathrm{d}^2}{\mathrm{d}r^2} & 0 \\
  0 & 0 & 0 & \frac{\mathrm{d}^2}{\mathrm{d}r^2} \\
\end{array}\right ).
\end{equation}
The topological charges appearing in (\ref{n4algbe1sjdjfgeneraldfgdfg}), can be classified more easily in comparison to the gapped graphene case, since the operator $\mathcal{D}_i$ is self-adjoint. Each of the topological charges $\mathcal{Z}_{ii }$ is given by,
\begin{equation}\label{newsusymat22}
\mathcal{Z}_{ii}=\left ( \begin{array}{ccccc}
  \mathcal{Z}^1_{ii} & 0 \\
  0 & \mathcal{Z}^2_{ii }  \\
\end{array}\right ).
\end{equation}
with the operator $\mathcal{Z}^1_{ii }$ being equal to,
\begin{equation}\label{newsusymat32}
\mathcal{Z}^1_{ii}=\left ( \begin{array}{cc}
-\frac{\mathrm{d}}{\mathrm{d}r}(\frac{l}{r})-\frac{l}{r}\frac{\mathrm{d}}{\mathrm{d}r}-\frac{l^2}{r^2}+\Delta_n^2(r) & (\frac{\mathrm{d}}{\mathrm{d}r}-\frac{l}{r})\Delta_n(r)+\Delta_n(r)(\frac{\mathrm{d}}{\mathrm{d}r}-\frac{m}{r}) \\
 (\frac{\mathrm{d}}{\mathrm{d}r}-\frac{m}{r})\Delta_n(r)+\Delta_n(r)(\frac{\mathrm{d}}{\mathrm{d}r}-\frac{l}{r}) & -\frac{\mathrm{d}}{\mathrm{d}r}(\frac{m}{r})-\frac{m}{r}\frac{\mathrm{d}}{\mathrm{d}r}-\frac{m^2}{r^2}+\Delta_n^2(r) \\
\end{array}\right )
\end{equation}
and with the operator $\mathcal{Z}^2_{ii }$ being equal to $\mathcal{Z}^1_{ii }$, that is $\mathcal{Z}^1_{ii }=\mathcal{Z}^2_{ii }$. In addition, the topological charges $\mathcal{Z}_{ij }$ are equal to,
\begin{equation}\label{newsusymat52}
\mathcal{Z}_{ i j }=\left ( \begin{array}{ccccc}
  \mathcal{Z}^1_{ i j } & 0 \\
  0 & \mathcal{Z}^2_{ i j }  \\
\end{array}\right ).
\end{equation}
with $\mathcal{Z}^1_{ i j }$ being equal to the matrix,
\begin{equation}\label{newsusymat62}
\mathcal{Z}^1_{i j }=\left ( \begin{array}{cc}
\frac{\mathrm{d}^2}{\mathrm{d}r^2}-\frac{\mathrm{d}}{\mathrm{d}r}(\frac{l'}{r})-\frac{l}{r}\frac{\mathrm{d}}{\mathrm{d}r}-\frac{ll'}{r^2}+\Delta_n(r)\Delta_{n'}(r) & (\frac{\mathrm{d}}{\mathrm{d}r}-\frac{l}{r})\Delta_{n'}(r)+\Delta_n(r)(\frac{\mathrm{d}}{\mathrm{d}r}-\frac{m'}{r}) \\
 (\frac{\mathrm{d}}{\mathrm{d}r}-\frac{m}{r})\Delta_{n'}(r)+\Delta_n(r)(\frac{\mathrm{d}}{\mathrm{d}r}-\frac{l'}{r}) & (\frac{\mathrm{d}}{\mathrm{d}r}-\frac{m}{r})(\frac{\mathrm{d}}{\mathrm{d}r}-\frac{m'}{r})+\Delta_n(r)\Delta_{n'}(r) \\
\end{array}\right )
\end{equation}
and the operator $\mathcal{Z}^1_{ i j }$, is given by,
\begin{equation}\label{newsusymat72}
\mathcal{Z}^2_{ i j }=\left ( \begin{array}{cc}
\frac{\mathrm{d}^2}{\mathrm{d}r^2}-\frac{\mathrm{d}}{\mathrm{d}r}(\frac{l}{r})-\frac{l'}{r}\frac{\mathrm{d}}{\mathrm{d}r}-\frac{ll'}{r^2}+\Delta_n(r)\Delta_{n'}(r) & (\frac{\mathrm{d}}{\mathrm{d}r}-\frac{l'}{r})\Delta_{n}(r)+\Delta_{n'}(r)(\frac{\mathrm{d}}{\mathrm{d}r}-\frac{m}{r}) \\
 (\frac{\mathrm{d}}{\mathrm{d}r}-\frac{m'}{r})\Delta_{n}(r)+\Delta_{n'}(r)(\frac{\mathrm{d}}{\mathrm{d}r}-\frac{l}{r}) & (\frac{\mathrm{d}}{\mathrm{d}r}-\frac{m'}{r})(\frac{\mathrm{d}}{\mathrm{d}r}-\frac{m}{r})+\Delta_n(r)\Delta_{n'}(r) \\
\end{array}\right )
\end{equation}
Therefore, given the number $n$, the electrons in superconducting graphene can form an extended $N=2n$ one dimensional supersymmetry with non trivial topological charges. Note that we have four sets of these $N=2n$ supersymmetries for the same reasons as explained in detail in the $N=2$ case.

\section*{Conclusions}

In this paper we studied some field theoretic attributes of two graphene configurations, namely in a gapped graphene setup and in superconducting graphene. Specifically, we found that the electron states constitute a number of one dimensional $N=2$ supersymmetries in both gapped and superconducting graphene. We explicitly demonstrated that these supersymmetries are unbroken for both cases. We have to note that there is no way of breaking these supersymmetries dynamically or spontaneously, since these are one dimensional supersymmetries and there is no way to achieve this. In the case of gapped graphene, the $N=2$ supersymmetries combine to form an $N=4$ one dimensional supersymmetry which has non-trivial topological charges. In the superconducting graphene case, the extended supersymmetric structure is much more involved and depends on the number of the electron zero modes around the vortex defect. If there exist $n$ distinct zero modes, then the extended supersymmetry is an $N=2n$ supersymmetry with non-trivial topological charges. In both cases these topological charges cannot be central charges due to the fact that these do not commute with all the operators of the algebra.

An interesting feature of both superconducting and gapped graphene is that the supersymmetries remain unbroken, a result that holds true even if hopping effects and compact perturbations of the gap function $\Delta (r)$ are taken into account. As we explicitly showed, the Witten index is robust against such kind of changes. In the case of superconducting graphene, in which case zero modes are considered, our result proves the validity of our findings, since these kind of changes never affect the Fredholm index of the associated to the system Dirac operators. Supersymmetry offers another point of view of the problem at hand. In addition, the same could apply for the gapped graphene, although the modes have a specific energy eigenvalue. The perspective of supersymmetry we adopted for gapped graphene is new and could possibly be an indicator of a non-linear underlying supersymmetry. The latter feature strongly validates the field theoretic limit of gapped graphene, which is also suggested and used in the relevant literature (see the review \cite{review}). We hope to further address the field theoretic character problem of gapped graphene in the future.

\end{document}